\documentclass[aps,prl,twocolumn]{revtex4}
\usepackage{epsfig}

\begin{document}

{\bf 
\noindent Comment on
``Jamming Percolation and Glass Transitions in Lattice
Models.''}

Toninelli et. al. recently introduced the knights
model, a correlated percolation system~\cite{TBF}. They
claimed to
prove that the
critical point of this model was the
same as that for directed percolation (DP), and
then used this 
to show (assuming a conjecture
about DP, described later in a more detailed proof~\cite{TB1})
that this model has a discontinuous
phase transition with
a diverging correlation length.
However, there is an error in 
their work, 
so that these results are unproven
for their model.
Their proofs can, however, be modified to work for a 
similar
model.

\begin{figure}[b]
\epsfig{figure=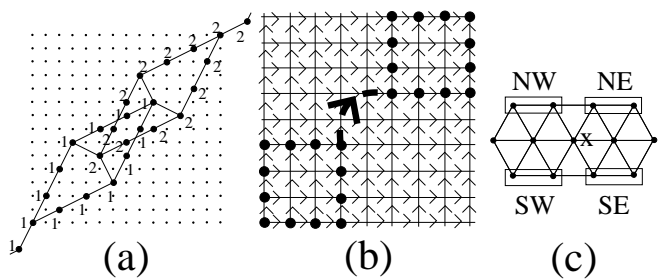,width=3.4in}
\caption{(a) Northeast chains in SL 1 end,
and are stabilized by northeast chains in 
SL 2. Numbers indicate sublattices.
(b) A configuration 
that receives an additional connection (the dashed line)
with
probability $s$.
(c) A modification of the knights model 
for which the proofs in~\cite{TBF} are valid.}
\label{fig:Sub1to2}
\label{fig:JumpPerc}
\label{fig:Newknights}
\end{figure}

The square lattice of the knights model can be
divided into three equivalent 
disjoint sublattices (SLs); one SL is
shown in Fig. 1a of~\cite{TBF}.
In their proof that 
$\rho_c^{\rm  knights}=\rho_c^{\rm DP}$,
Toninelli et. al. 
claim that
a site at the corner of a very large void can only be stable 
if it is part of a very long
DP
path in a single SL
(see Fig. 1c of~\cite{TBF}). 
However, this claim is not true. 
As shown in Fig.~\ref{fig:Sub1to2}a,
a northeast directed chain in
one SL (\#1) can terminate in a diamond,
and be
stabilized by a diamond in a different
SL (\#2). 
A site at the corner of a void can be made stable
by a
series of such linked diamonds, cycling between the
three SLs, and never having DP
paths of length greater than 7 in any SL.
The claim in~\cite{TBF}
is thus false, and the subsequent results that rely on it,
such as the discontinuity of the transition and the diverging
correlation length, are currently unproven.

Is it possible that despite this error,
$\rho_c^{\rm knights}=\rho_c^{\rm DP}$?
The following argument shows this to be unlikely.
Consider a modification of DP that we call
Jumping DP.
As in DP, we occupy sites on the square lattice
with probability $p$, and have
directed bonds to the north and east.
However, we define 
additional connections. We divide the
lattice into disjoint 9x9 blocks, and for each block if the
two hollow squares of sites shown in
Fig.~\ref{fig:JumpPerc}b
are occupied,
we with probability $s$ add
a directed bond from the southwest square to the
northeast square.
The critical point of this model is a
function of $s$: $\rho_c^{\rm Jump}(s)$.

By repeating Fig.~\ref{fig:Sub1to2}a three times, to create diamonds
connecting SL \#1 to \#2 to \#3 to
\#1, we obtain a structure that 
links two separated 
diamonds in SL \#1, through 24 sites in the other two
SLs. The mapping from SL \#1 to the more
conventional form of DP takes the
occupied sites of SL \#1 
in this structure
to the marked sites in 
Fig.~\ref{fig:JumpPerc}b. So
if we restrict ourselves to looking at 
sites in SL \#1, sites that appear disconnected
may be connected by these 24 sites.
Infinite chains in Jumping DP
are stable infinite clusters in the knights
model, and thus
$\rho_c^{\rm knights} \leq \rho_c^{\rm Jump}(p^{24}) 
\leq \rho_c^{\rm DP}$.

It is reasonable to think that these extra connections
should depress the critical probability, so that
$\rho_c^{\rm Jump}(s)<\rho_c^{\rm DP}$ for all $s>0$, implying
$\rho_c^{\rm knights}<\rho_c^{\rm DP}$. 
In~\cite{Enhancement}
it was shown that if percolation 
was ``enhanced'' by adding, for specified subconfigurations,
extra local connections 
with probability $s$, this would
{\it strictly} decrease the critical probability, for 
any $s>0$, so long as 
a single enhancement could create a doubly-infinite
path where none existed before.
The results of~\cite{Enhancement}
are for undirected percolation, and thus
do not provide a mathematically rigorous proof here.
But they are analogous enough to strongly suggest that
$\rho_c^{\rm Jump}(s)<\rho_c^{\rm DP}$ for all $s>0$.
It is difficult to see how adding such a new route for
paths to infinity could leave the critical probabiltiy 
completely unchanged.

The situation in Fig.~\ref{fig:Sub1to2}a
can occur only because 
in the knights model two 
northeast chains can cross 
without having sites in common.
If the knights model is modified
to one where this is impossible,
as with the neighbor connections on the
triangular lattice shown in
Fig.~\ref{fig:Newknights}c (to be compared with Fig. 1b
of~\cite{TBF}), 
the results 
found in~\cite{TBF} hold, and
the critical point is $\rho_c^{\rm DP}$.
One complication is that the proofs
in~\cite{TBF} 
that the transition is
discontinuous with a diverging correlation
length implicitly 
assume that if northeast and northwest
chains cross, they must have sites in
common, so that one path blocks the other,
but this assumption is neither true
for the knights model, nor for the model in
Fig.~\ref{fig:Newknights}c.
However, the proofs can be modified
so that everywhere northeast and northwest chains
cross, they cross in a large number of places.
It can then be shown that 
since a local change in the vicinity of any crossing
can create the desired blocking, at least one such
blocking occurs with high probability. The proofs
in~\cite{TBF} can thus be made valid 
(assuming the conjecture in~\cite{TB1})
for the model in Fig.~\ref{fig:Newknights}c,
with one exception:
we are unable to verify the relatively
tight upper bound on the crossover length 
claimed
in~\cite{TBF}.\\

\noindent M. Jeng and J. M. Schwarz \\
Physics Dept., Syracuse University, Syracuse, NY 13244 


\end{document}